\documentclass[aps,prapplied,reprint,superscriptaddress,longbibliography]{revtex4-2}

\usepackage[T1]{fontenc}
\usepackage[utf8]{inputenc}
\usepackage{amsmath,amssymb,mathtools,bm}
\usepackage{graphicx}
\usepackage{xcolor}
\usepackage[colorlinks=true, linkcolor=black, citecolor=black, urlcolor=black]{hyperref}

\newcommand{\en}{\text{-- }}

\begin{document}

\title{The critical role of substrates in mitigating the power-efficiency trade-off \\ in near-field thermophotovoltaics}

\author{Kartika N. Nimje}
\affiliation{ICFO – Institut de Ciències Fotòniques, The Barcelona Institute of Science and Technology, Mediterranean Technology Park, Av. Carl Friedrich Gauss 3, Castelldefels (Barcelona) 08860, Spain}

\author{Julien Legendre}
\affiliation{ICFO – Institut de Ciències Fotòniques, The Barcelona Institute of Science and Technology, Mediterranean Technology Park, Av. Carl Friedrich Gauss 3, Castelldefels (Barcelona) 08860, Spain}

\author{Michela F. Picardi}
\affiliation{ICFO – Institut de Ciències Fotòniques, The Barcelona Institute of Science and Technology, Mediterranean Technology Park, Av. Carl Friedrich Gauss 3, Castelldefels (Barcelona) 08860, Spain}

\author{Alejandro W. Rodriguez}
\affiliation{Department of Electrical and Computer Engineering, Princeton University, Princeton, NJ 08544, USA}

\author{Georgia T. Papadakis}
\email{georgia.papadakis@icfo.eu}
\affiliation{ICFO – Institut de Ciències Fotòniques, The Barcelona Institute of Science and Technology, Mediterranean Technology Park, Av. Carl Friedrich Gauss 3, Castelldefels (Barcelona) 08860, Spain}


\begin{abstract}
Near-field thermophotovoltaic systems can achieve ultra-high power densities, however, this often comes at the cost of reduced efficiency. We show that this power–efficiency trade-off can be mitigated through substrate engineering. We exploit gradient-based optimization and show that thin lossless metallic films with plasma frequencies resonantly matched to the plasmonic emitter can yield high power and spectral efficiency by spectrally enhancing and confining radiative heat transfer to a narrow spectral range just above the photovoltaic bandgap. Compared to noble metals and air-bridged structures, designs deriving from such optimization yield more than an order-of-magnitude increase in radiative power density while maintaining high efficiency. Our results highlight the critical role of the substrate and the potential of substrate optimization for overcoming fundamental limitations of near-field thermophotovoltaic systems.
\end{abstract}

\maketitle

\par{Thermophotovoltaic (TPV) systems convert thermal radiation into electrical power. They are increasingly being explored for applications such as industrial waste-heat recovery, space-based energy generation, and thermal energy storage \cite{datas_thermophotovoltaic_2021}. In these systems, a hot emitter radiates energy towards a photovoltaic (PV) cell across a vacuum gap. At large emitter–cell separation distances (far-field regime), this exchange is bounded by the blackbody limit. However, at sub-micron gaps (near-field regime), photons tunnel across the gap, enabling radiative heat transfer that surpasses far-field bounds by orders of magnitude \cite{polder_theory_1971,pendry_radiative_1999}. Surface polariton resonances in the emitter can further enhance this transfer, enabling near-field TPVs to achieve extremely high radiative power densities \cite{narayanaswamy_surface_2003, laroche_near-field_2006, shen_surface_2009, mittapally_near-field_2021,lucchesi_near-field_2021}.

\par{Despite this promise, near-field TPVs face a longstanding challenge common to all radiative heat engines \cite{giteau_thermodynamic_2023}: a fundamental trade-off between power output and conversion efficiency. While increasing emitter-cell coupling boosts total radiative flux, it also amplifies spectral components that cannot be converted to electricity; photons with energies below the PV cell’s bandgap contribute to parasitic heating, while photons far above the bandgap produce carriers that rapidly thermalize. In far-field TPVs, this power-efficiency trade-off has been addressed through spectral engineering, such as selective emitters \cite{nefzaoui_selective_2012, xiao_ultrabroadband_2025}, multi-junction PV cells \cite{datas_optimum_2015, king_fundamental_2025}, hot-carrier TPVs \cite{nimje_hot-carrier_2024}, and reflective mirrors beneath the PV cell that recycle sub-bandgap photons \cite{omair_ultraefficient_2019,  fan_near-perfect_2020}. These strategies have enabled far-field systems to achieve conversion efficiencies exceeding 40\% in practice \cite{lapotin_thermophotovoltaic_2022}. However, such techniques cannot be directly extended to the near-field regime, which is governed by evanescent mode coupling.}

\par{Efforts to mitigate parasitic losses in near-field TPVs have primarily focused on shaping the emitter spectrum \cite{karalis_squeezing_2016} or inserting intermediate filtering layers \cite{inoue_near-field_2018}. Yet, one key component remains overlooked: the PV cell’s substrate. In the near field, the substrate is not a passive reflector but an active participant in the radiative exchange, as it can couple directly with the emitter \cite{bright_performance_2014}. The optical properties of the substrate thus critically determine whether evanescent modes contribute constructively to energy conversion or lead to parasitic losses. A poorly designed substrate can absorb a substantial portion of the heat flux, severely degrading efficiency. Recent efforts have considered air-bridge designs in near-field TPV systems \cite{inoue_near-field_2021,feng_improved_2022,lee_air-bridge_2022,roy-layinde_sustaining_2022,lim_enhanced_2023}, as these indeed partially mitigate substrate absorption. However, such solutions often fail to fully optimize the spectral selectivity in the near-field regime. In fact, what constitutes an optimal substrate for performance enhancement in near-field TPVs remains an open question.}

\par{In this work, we show that substrate engineering, even in planar geometries, offers a compelling strategy for alleviating the power-efficiency trade-off in near-field TPVs. Using fluctuational electrodynamics combined with nonlinear gradient-based optimization, we identify an unconventional yet effective solution: a thin lossless metallic (Drude) substrate with a carefully tailored plasma frequency. Such a substrate allows resonant coupling between the surface plasmon polaritons (SPPs) supported at the emitter-vacuum and cell-substrate interfaces (see Fig. \ref{fig:fig1}). This coupling maximizes photon absorption at frequencies near the bandgap, thereby enhancing the radiative power delivered to the cell. Simultaneously, it suppresses both below-bandgap absorption as well as thermalization by confining photon tunneling to a narrow spectral band, which enables high efficiency.}

\begin{figure}
    \centering
    \includegraphics[width=\linewidth]{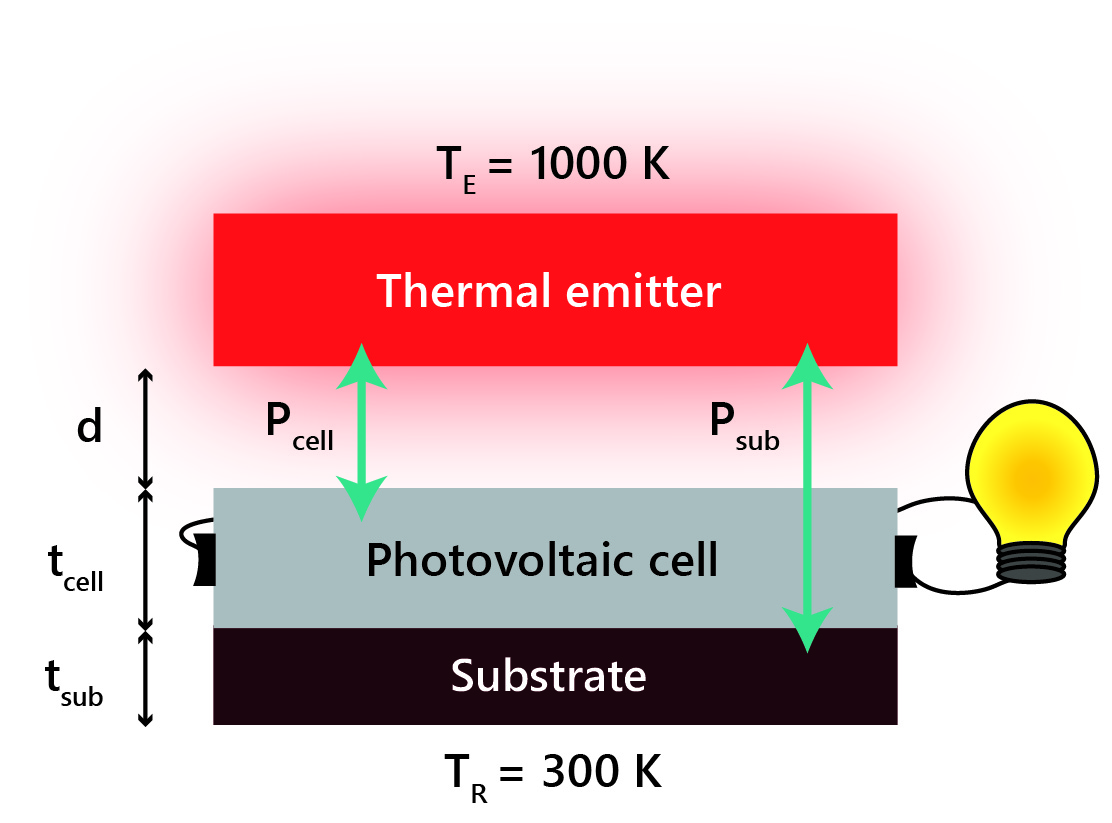}
    \caption{Schematic of the near-field TPV system comprising a semi-infinite thermal emitter at temperature $T_\mathrm{E} = 1000 \ \mathrm{K}$, a vacuum gap $d = 10 \ \mathrm{nm}$, a photovoltaic cell of thickness $t_\mathrm{cell}$, and a substrate of thickness $t_\mathrm{sub}$, both held at $T_\mathrm{R} = 300 \ \mathrm{K}$. $P_\mathrm{cell}$ denotes the heat flux exchanged between the emitter and the cell, while $P_\mathrm{sub}$ denotes the flux exchanged between the emitter and the substrate, including radiative leakage beyond the substrate. For below-bandgap frequencies, the cell remains transparent, allowing direct radiative coupling between the emitter and the substrate.}
    \label{fig:fig1}
\end{figure}

\par{We consider a near-field TPV system consisting of a semi-infinite plasmonic emitter made of indium tin oxide (ITO), held at $T_\mathrm{E} = 1000$ K, as illustrated in Fig. \ref{fig:fig1}. Conductive oxides such as ITO support tunable plasmonic resonances at frequencies close to the bandgap of typical TPV cells, yielding improved near-field TPV performance. The dielectric response of ITO is described by a Drude model:
\begin{equation}
\label{eq: emitter}
\varepsilon_\mathrm{em}(\omega) = \varepsilon_{\infty,\mathrm{em}} \left[1 - \frac{\omega_\mathrm{em}^2}{\omega(\omega + i \gamma_\mathrm{em})} \right],
\end{equation}
where $\varepsilon_{\infty,\mathrm{em}} = 4$, $\omega_\mathrm{em} = 0.45$ eV, and $\gamma_\mathrm{em} = 0.1$ eV \cite{zhao_high-performance_2017}. A vacuum gap of $d = 10$ nm separates the emitter from the photovoltaic cell, ensuring operation in the near-field, which enables strong heat transfer between the emitter and cell. In particular, the power enhancement at this vacuum gap exceeds the blackbody limit by a factor of 40 \cite{fiorino_nanogap_2018}. To simplify the analysis, nonlocal optical effects are neglected.}

\par{The PV cell, which is placed atop a substrate (Fig. \ref{fig:fig1}), is maintained at a uniform temperature of $T_\mathrm{R} = 300$ K. The cell has a thickness $t_\mathrm{cell}$, while the substrate beneath it has a thickness $t_\mathrm{sub}$. The dielectric response of the cell is modeled as:
\begin{equation}
\label{eq: cell}
\varepsilon_\mathrm{cell}(\omega) = \left( n_\mathrm{cell} + i \frac{\alpha_\mathrm{cell}}{2k_0} \right)^2, 
\end{equation}
where $n_\mathrm{cell}$ is the refractive index, $k_0 = \omega /\mathrm{c}$ is the free-space wavenumber, and $\alpha_\mathrm{cell}$ is the absorption coefficient. Following standard semiconductor behavior \cite{ilic_overcoming_2012}, $\alpha_\mathrm{cell}$ exhibits a square-root dependence above the bandgap:
\[ \alpha_\mathrm{cell}(\omega) = 
\begin{cases} 
0, & \omega < \omega_g \\
\alpha_\mathrm{0,cell} \sqrt{\frac{\omega - \omega_g}{\omega_g}}, & \omega \geq \omega_g
\end{cases}.\]
}

\par{We consider indium arsenide (InAs) as the PV cell material, with refractive index $n_\mathrm{cell} = 3.51$, bandgap frequency $\omega_g = 0.36$ eV, and absorption coefficient $\alpha_{0,\mathrm{cell}} = 1.3 \times 10^6 \ \text{m}^{-1}$ \cite{ilic_overcoming_2012}. Below the bandgap ($\omega < \omega_g$), the cell is effectively transparent, exhibiting zero absorption. This is reflected in the step-like imaginary part of its dielectric function, as shown in Supporting Information (SI). To ensure strong near-field coupling between the emitter and the cell, the emitter's plasma frequency $\omega_\mathrm{em}$ is chosen such that the surface plasmon resonance of ITO, given by $\omega_\mathrm{res,em} = \omega_\mathrm{em} \sqrt{\varepsilon_{\infty,\mathrm{em}} / (\varepsilon_{\infty,\mathrm{em}} + 1)} \simeq 0.4$ eV, lies slightly above the bandgap ($\omega_\mathrm{res,em} > \omega_g$) \cite{zhao_high-performance_2017}.
}

\begin{figure*}
    \centering
    \includegraphics[width=1\linewidth]{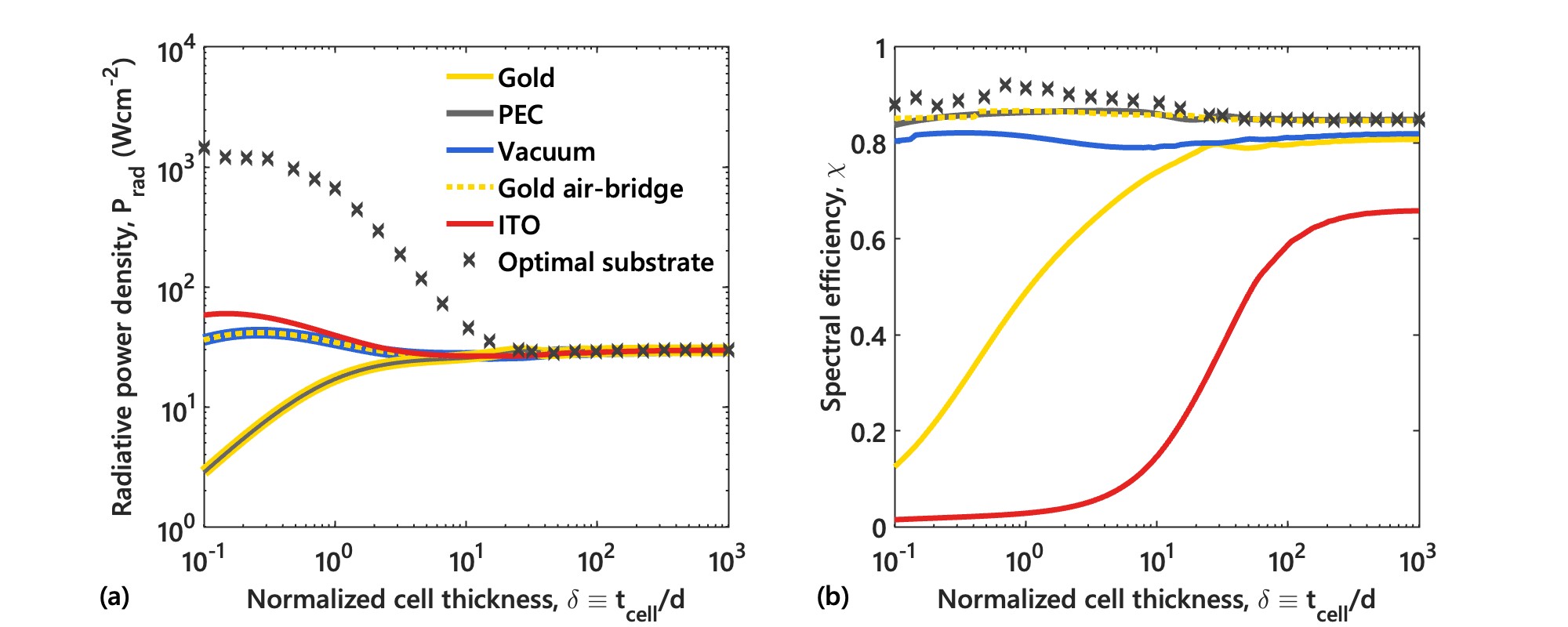}
    \caption{Cell thickness as a key parameter in near-field interactions: (a) Radiative power available for optoelectronic conversion, $P_\mathrm{rad}$ (in $\mathrm{Wcm^{-2}}$), and (b) spectral efficiency, $\chi$, plotted as functions of normalized cell thickness, $\delta$ with respect to the 10 nm vacuum gap, for various substrates. Discrete markers (gray crosses) indicate optimal values obtained by jointly optimizing material parameters and substrate thickness for single-layer substrates. For completeness, the total parasitic heat flux lost in the cell and substrate ($P_\mathrm{lost}$) is provided in the Supporting Information.}
    \label{fig:fig2}
\end{figure*}

\par{To understand the role of the substrate in near-field radiative heat transfer, we focus on two key photonic quantities: the radiative power above the PV cell’s bandgap ($P_\mathrm{rad}$) and the spectral efficiency ($\chi$). The quantity $P_\mathrm{rad}$ represents the portion of the spectrum available for optoelectronic conversion in the cell, while $\chi$ captures how effectively the system manages the incoming photon spectrum, accounting for photon recycling and losses due to sub-bandgap absorption and carrier thermalization \cite{burger_present_2020}. In the limit of unity internal quantum efficiency, these quantities reduce to the conventional electrical figures of merit: the output electrical power density ($P_\mathrm{el}$) and the conversion efficiency ($\eta$). The quantity $\chi$ is defined as:
\begin{equation}
\label{eq: chi}
    \chi \equiv \frac{P_\mathrm{rad}}{P_\mathrm{tot}},
\end{equation}
where $P_\mathrm{rad}$ and $P_\mathrm{tot}$ are given by:
\begin{equation}
\label{eq: prad}
    P_\mathrm{rad} = \int_{\omega_g}^{\infty} \int_0^{\infty} \hbar \omega_\mathrm{g} \ \zeta_\mathrm{cell} \, d \beta d \omega,
\end{equation}

\begin{equation}
\label{eq: ptot}
    P_\mathrm{tot} = \int_0^{\infty} \int_0^{\infty} \hbar \omega \ (\zeta_\mathrm{cell}+ \zeta_\mathrm{sub})  \, d \beta d \omega = P_\mathrm{cell} + P_\mathrm{sub}.
\end{equation}
 $P_\mathrm{tot}$ is the net photonic heat flux exchanged between the emitter and the combination of PV cell and substrate ($P_\mathrm{cell}$ + $P_\mathrm{sub}$), encompassing the useful power $P_\mathrm{rad}$, as well as thermalization losses within the cell, parasitic absorption in the substrate, and radiative leakage beyond the substrate. In the above equation, the quantity $\zeta_\mathrm{cell/sub}$ is given by: 
\begin{equation}
\label{eq: zeta_i}
    \zeta_\mathrm{cell/sub} = \dfrac{\beta \omega^2}{4  \mathrm{\pi^2} \mathrm{c^2}}  [\Phi(\omega, T_\mathrm{E}) - \Phi(\omega, T_\mathrm{R})] \xi_\mathrm{cell/sub}\left(\beta, \omega \right),
\end{equation}
where $\xi_\mathrm{cell/sub}\left(\beta, \omega \right)$ denotes the photon tunneling probability \cite{song_near-field_2015} at frequency $\omega$ and normalized in-plane wavevector $\beta$, and $\Phi\left( \omega, T \right) = \left[\exp \left(\hbar \omega / k_\mathrm{B} T\right) - 1\right]^{-1}$ is the Bose-Einstein occupation number. In  SI, we analytically express $\xi_\mathrm{sub}$ for a semi-infinite substrate for both propagating and evanescent modes.}

\par{We first consider conventional substrates in TPVs \en gold, a perfect electric conductor (PEC), vacuum (no substrate), a gold mirror separated by a $1\,\mu\mathrm{m}$ air gap (air-bridge), and an ITO substrate. All aforementioned substrates couple differently with the ITO emitter. It is known that the thickness of the TPV cell dramatically affects device performance in the near field \cite{papadakis_thermodynamics_2021}. In addition to influencing the cell’s optical opacity, the cell thickness determines the distance between the emitter and the substrate, and thus affects their coupling. Based on this, three regimes can be identified in Fig. \ref{fig:fig2}: a thin-cell regime (normalized cell thickness $\delta < 30$, where $\delta = t_{\mathrm{cell}}/d$) in which the emitter and substrate are close enough to exchange radiation across the entire spectrum (both above and below the bandgap); a thick-cell regime ($30 < \delta < 300$) where the cell is optically thick to above-bandgap radiation and surface modes coupling can only occur below the bandgap, causing the radiative power $P_{\mathrm{rad}}$ delivered to the cell to plateau; and an ultra-thick regime ($\delta > 300$) where the substrate is effectively in the far field and only propagating and frustrated modes contribute to heat transfer, causing both $P_{\mathrm{rad}}$ and the spectral efficiency $\chi$ to asymptotically saturate. These considerations indicate that substrate properties are most consequential in the thin-cell limit, where near-field coupling is strongest.}

\par{Fig. \ref{fig:fig2}(a) compares the radiative power density $P_{\mathrm{rad}}$ for the different substrates as a function of normalized cell thickness $\delta$. For gold and PEC substrates, which do not couple strongly with the ITO emitter, $P_{\mathrm{rad}}$ is relatively low in the thin-cell regime. As the cell thickens,  $P_{\mathrm{rad}}$ increases; as expected, a thicker cell simply absorbs more radiation. By contrast, with an ITO substrate or even a vacuum or air-bridge backing, $P_{\mathrm{rad}}$ can be enhanced for thinner cells due to improved near-field coupling. In fact, the highest power is obtained using a semi-infinite ITO substrate in the ultrathin limit, since the plasmonic ITO emitter can resonantly couple to an identical ITO on the backside at frequencies near the cell bandgap, greatly boosting above-bandgap radiative transfer. The vacuum and air-bridge cases also yield higher $P_{\mathrm{rad}}$ at small $\delta$ compared to gold or PEC. In these configurations, evanescent waves can tunnel between the two interfaces of the thin cell (emitter side and backside), increasing absorption in the cell even without a directly adjacent substrate \cite{francoeur_near-field_2008}. However, because vacuum and the distant gold mirror largely suppress direct emitter–substrate coupling, their $P_{\mathrm{rad}}$ in Fig. \ref{fig:fig2}(a) remains a bit lower than the ITO–ITO pairing. Overall, in the thin-cell regime we see that introducing some coupling (even in absence of an absorbing substrate) can improve cell absorption compared to an ideal mirror or a poorly matched metal. As the cell becomes optically thick, all substrates eventually converge, making the choice of substrate largely irrelevant.}

\par{Fig. \ref{fig:fig2}(b) shows the corresponding spectral efficiency $\chi$. Plasmonic materials like ITO and gold, are detrimental to $\chi$ in the near-field regime \en strong evanescent coupling enables significant radiative exchange below the bandgap (where the cell is transparent), leading to parasitic absorption in the substrate and a low spectral efficiency. This effect is most pronounced for thin cells ($\delta < 30$), for which achieving high efficiency requires minimizing sub-bandgap coupling. Indeed, lossless substrates such as vacuum yield much higher $\chi$ at small $\delta$ by eliminating below-bandgap heat transfer due to evanescent modes. A PEC mirror represents the ideal limit, completely reflecting all photons for potential recycling, and thus maintaining a high $\chi$. Notably, the gold air-bridge approach achieves efficiency almost on par with the PEC case. By placing the gold mirror $1\,\mu\mathrm{m}$ away, sub-bandgap photon tunneling into the substrate is prevented (similar to vacuum) while far-field (propagating) photons that would otherwise escape the system are reflected back, yielding a high $\chi$ without sacrificing much radiative power. Still, none of these conventional substrates can achieve both high power and high efficiency simultaneously.}

\par{To overcome this power-efficiency trade-off, we next turn to optimizing the substrate, employing nonlinear gradient-based local optimization tailored to the nonconvex nature of near-field radiative transfer \cite{jin_overcoming_2017, zhang_optimal_2020}. The search space spans both material parameters and substrate thicknesses, subject to physical constraints (for details on optimization, see SI). An ideal substrate should increase the radiative power delivered to the PV cell while suppressing parasitic absorption in the substrate. Accordingly, we define the objective function as
\begin{equation}
\label{eq: objectivefn}
\Psi = P_\mathrm{rad} - P_\mathrm{sub}.
\end{equation}
To allow for a wide range of potential solutions, we began our optimization considering a general model encompassing semiconductors (Eq. \ref{eq: cell}), plasmonic (Eq. \ref{eq: emitter}), and phononic (Lorentz \cite{caldwell_low-loss_2015}) materials as the dielectric function of the substrate, and search for solutions that maximize $\Psi$ (Eq. \ref{eq: objectivefn}). Interestingly, for thin cells, the optimization consistently yield a Drude-like response, labeled $\varepsilon_\mathrm{sub}(\omega, \varepsilon_{\infty, \mathrm{sub}}, \omega_\mathrm{sub}, \gamma_{\mathrm{sub}})$ henceforth. This natural selection can be understood via elimination; Lorentz-type dielectric functions with realistic phonon lifetimes \cite{caldwell_low-loss_2015} are too narrow to provide sufficient above-bandgap radiative heat transfer, and doped semiconductors offer spectrally broader yet weaker evanescent coupling, making both suboptimal. For thicker cells ($\delta \gtrsim 30$), multiple optima begin to emerge, with PEC becoming one viable solution. The gray crosses in Fig. \ref{fig:fig2} denote the outcomes of substrate optimization performed for discrete cell thicknesses. Each marker corresponds to the optimal values obtained by jointly optimizing material parameters and substrate thickness. These optimized configurations yield substantial improvements over all conventional substrates discussed previously. For example, at $\delta = 1$ (corresponding to a 10 nm cell), the optimal parameters are $\varepsilon_{\infty, \mathrm{sub}} = 1$, $\omega_\mathrm{sub} = 1.33$ eV, $\gamma_\mathrm{sub} = 0$ eV, and $t_\mathrm{sub} = 15.4$ nm (the variation of these parameters with cell thickness is shown in SI). These yield $P_\mathrm{rad} = 660$ Wcm$^{-2}$ and $\chi = 0.914$. This corresponds to a 40-fold enhancement in radiative power density and a 32-fold increase in spectral efficiency compared to a gold substrate. Additionally, the optimal substrate leads to a 1.05-fold enhancement in spectral efficiency compared to gold air-bridge substrate, and a 17-fold increase in radiative power compared to ITO substrate. Notably, for thin cells, these values exceed those achieved with an ideal PEC reflector.}

\begin{figure*}
    \centering
    \includegraphics[width=0.95\textwidth]{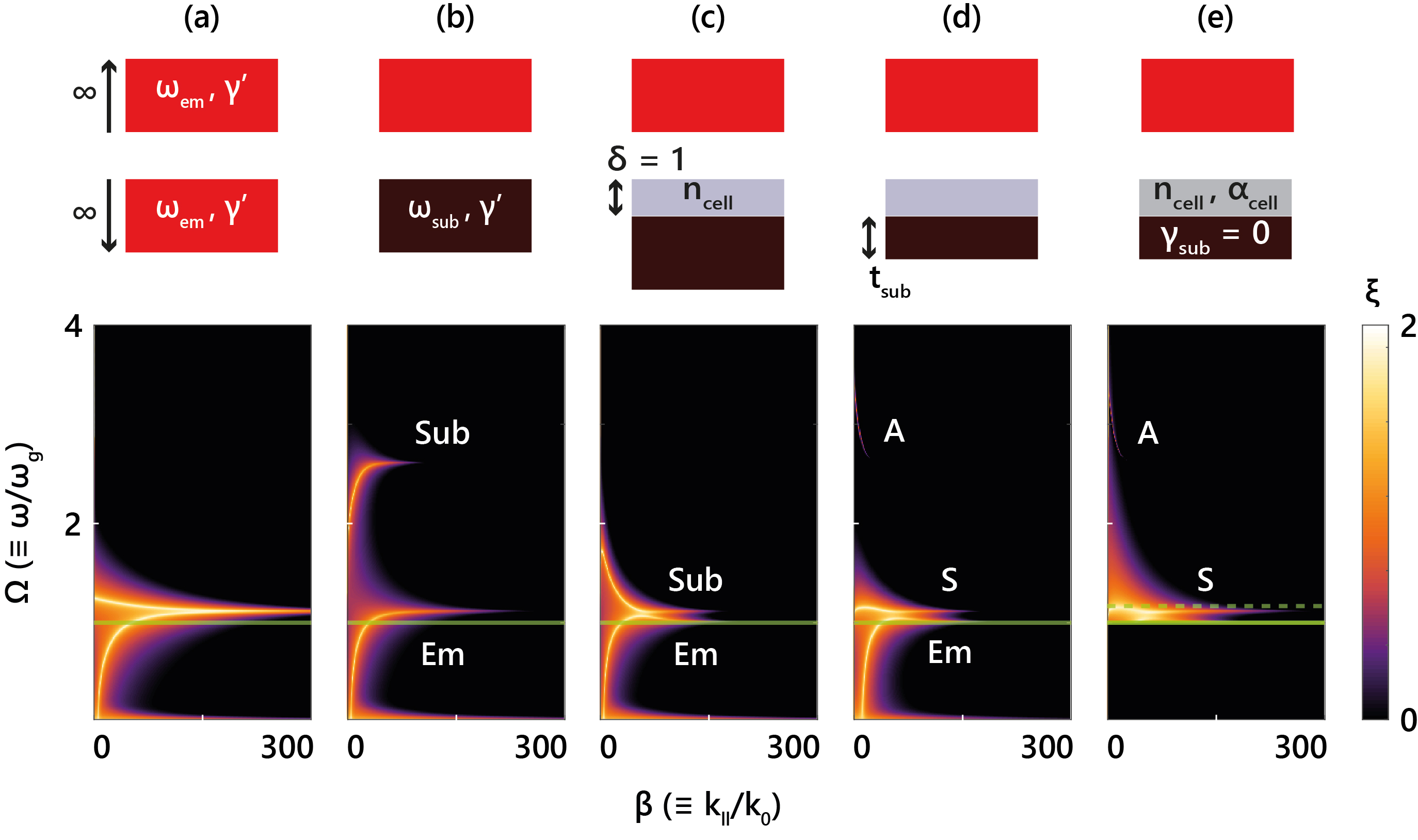}
    \caption{Evolution of photon tunneling: Top: schematic configurations; bottom: corresponding photon tunneling probability ($\xi_\mathrm{cell} + \xi_\mathrm{sub}$) versus normalized frequency ($\Omega \equiv \omega/\omega_\mathrm{g}$) and in-plane wavevector ($\beta \equiv k_\parallel/k_0$). The green solid line marks $\Omega = 1$ (bandgap frequency). (a) Identical semi-infinite ITO layers support degenerate SPPs. (b) Increasing the plasma frequency of the substrate yields an emitter- (\texttt{Em}) and a substrate-mode (\texttt{Sub}). (c) Adding a 10 nm $(\delta = 1)$ dielectric redshifts the \texttt{Sub} mode above the bandgap, enabling spectral alignment with \texttt{Em} mode. (d) Reducing substrate thickness induces hybridization into symmetric (\texttt{S}) and antisymmetric (\texttt{A}) modes. (e) A lossless substrate suppresses sub-bandgap tunneling, enhancing spectral selectivity. Replacing the dielectric layer with the cell broadens the above-bandgap absorption in the cell. The green dashed line denotes the narrow bandwidth of the spectral heat flux. }
    \label{fig:fig3}
\end{figure*}

\par{To explain the enhancement in near-field TPV performance enabled by the optimized substrate, we analyze the photon tunneling probability, $\xi(\omega,\beta)$, starting from a toy model of a symmetric emitter-to-emitter system, as shown in Fig. \ref{fig:fig3}a. Two identical semi-infinite ITO layers form a canonical metal–insulator–metal (MIM) geometry \cite{maier_plasmonics_2007}, supporting bonding and antibonding SPP modes. For visual clarity, the ITO loss is reduced to $\gamma’ = 5$ meV. The degenerate modes exhibit maximal photon tunneling probability \texttt{Em} at the ITO's SPP resonance ($\omega_\mathrm{res,em}$), enabling the strongest near-field coupling \cite{pascale_tight_2023}. Replacing the bottom ITO layer with a plasmonic metal, termed ``substrate'' henceforth, of higher plasma frequency ($\omega_\mathrm{sub} = 1.33$ eV) lifts this degeneracy. As shown in Fig. \ref{fig:fig3}b, \texttt{Em} exhibits two distinct resonant branches \en \texttt{Em} and \texttt{Sub} \en associated with the respective SPPs of the emitter and substrate.} 

\par{When a 10 nm ($\delta = 1$) dielectric layer is introduced above the substrate (Fig. \ref{fig:fig3}c), the resonance condition for the excitation of the \texttt{Sub} mode, now localized at the dielectric/substrate interface, shifts to lower frequencies. The \texttt{Em} mode is not affected by the dielectric layer significantly, since it is located at the air/emitter interface. It is ideal to redshift the \texttt{Sub} mode to frequencies just above the cell's bandgap, $\omega_g$, allowing it to couple with the \texttt{Em} mode for maximal heat transfer. To achieve this, the refractive index of the dielectric layer plays a key role; empirically, one can write \cite{economou_surface_1969}: 
\begin{equation}
\label{eq: resonance_condition}
\omega_\mathrm{sub}\sqrt{\dfrac{\varepsilon_{\infty,\mathrm{sub}}}{\varepsilon_{\infty,\mathrm{sub}} + n_\mathrm{cell}^2}} \simeq \omega_\mathrm{em} \sqrt{\dfrac{\varepsilon_{\infty,\mathrm{em}}}{\varepsilon_{\infty,\mathrm{em}} + 1}} \gtrsim \omega_g.
\end{equation}
When this condition is met, photon tunneling is spectrally concentrated near a common resonance just above $\omega_g$, maximizing useful energy transfer. The selection of a substrate with a plasma frequency larger than that of ITO warrants the proper alignment between the \texttt{Em} and \texttt{Sub} mode in the presence of the dielectric layer, which, as discussed below, represents the thin PV cell of the TPV configuration.}

\par{Next, we reduce the substrate thickness in Fig. \ref{fig:fig3}(c) to a finite, optimized value ($t_\mathrm{sub} = 15.4$ nm). This leads to mode hybridization of the \texttt{Sub} mode into symmetric (\texttt{S}) and antisymmetric (\texttt{A}) branches, as shown in Fig. \ref{fig:fig3}(d). The \texttt{A} mode, lying at a higher frequency, contributes negligibly to radiative exchange due to its spectral misalignment with the Planck distribution at $T_\mathrm{E}$. As a result, the above-bandgap heat transfer, primarily mediated by the \texttt{S} mode, becomes confined to frequencies just a few $k_\mathrm{B} T_\mathrm{E}$ above the bandgap (as shown in SI), which leads to high efficiency \cite{mcsherry_effects_2019, papadakis_broadening_2020}. Contributions from the \texttt{Em} mode below the bandgap can be effectively eliminated by reducing the losses of the substrate to zero (Fig. \ref{fig:fig3}e), thus eliminating any heat flux at $\omega<\omega_\mathrm{g}$. This analysis is independent of the presence of loss in the cell, which merely broadens the emission above the bandgap, as shown in Fig. \ref{fig:fig3}e. These observations clarify why the optimal near-field TPV substrate is a lossless, thin plasmonic film with a plasma frequency higher than that of the emitter. As shown conceptually in Fig. \ref{fig:fig3}, this selection of substrate maximizes emitter-substrate coupling in the presence of the cell. This, in turn, also maximizes absorption in the cell ($\xi_\mathrm{cell}$ is shown in SI), yielding large $P_\mathrm{rad}$ while suppressing below-bandgap absorption and thermalization losses, enabling $\chi$ higher than that for a PEC.}

\par{One may also consider the degree to which additional substrate complexity contributes to performance. We first confine the analysis to a three-layer configuration comprising the optimized plasmonic substrate, a vacuum spacer, and a far-field mirror. This design aims to recycle any residual propagating modes of radiation that leak out of the substrate. However, the performance gain is small \en less than $1\%$ \en since evanescent modes dominate the heat transfer. We also examine a carefully optimized bilayer, which similarly improves the objective function (Eq. \ref{eq: objectivefn}) by less than $1\%$. These results confirm that once the surface plasma frequency of the substrate is properly tuned, additional structural complexity yields diminishing returns. A single, lossless plasmonic layer suffices to achieve near-optimal performance.}

\begin{figure}
    \centering
    \includegraphics[width=\linewidth]{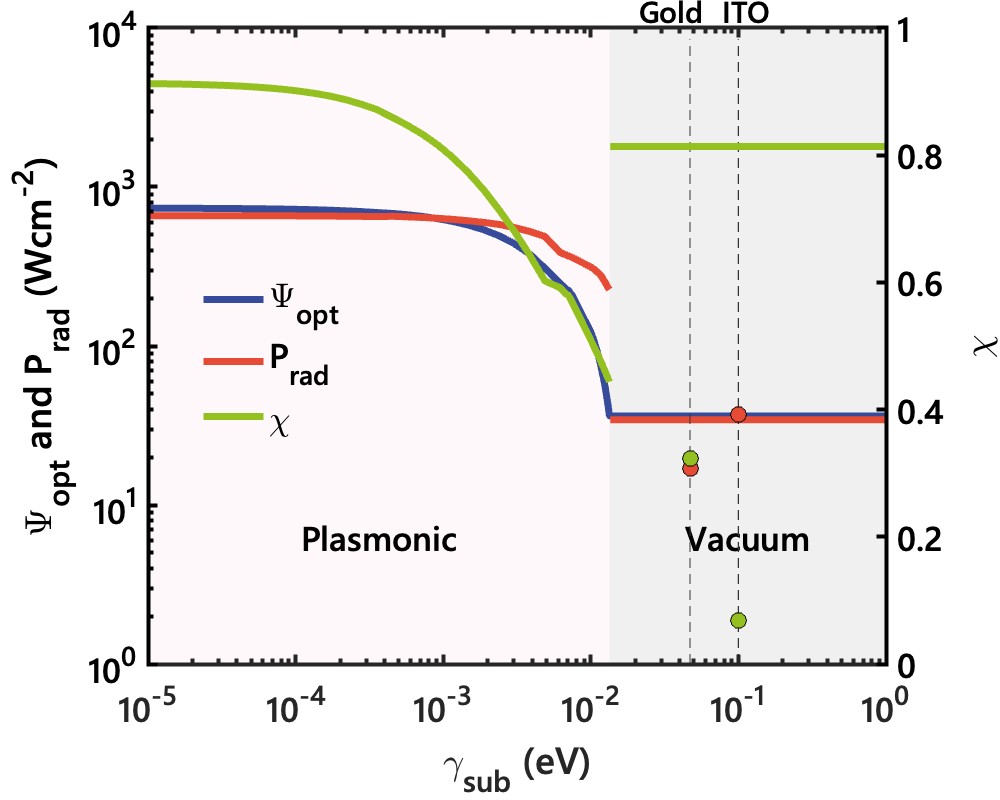}
    \caption{Objective function, $\Psi_\mathrm{opt}$ (blue), and corresponding radiative power density, $P_\mathrm{rad}$ (red), in units of $\mathrm{Wcm^{-2}}$, as functions of material loss $\gamma_{\mathrm{sub}}$ expressed in eV. The green curve indicates the associated spectral efficiency, $\chi$. The pink-shaded region corresponds to regimes where the optimal substrate is plasmonic, whereas the gray-shaded region indicates that vacuum is the optimal substrate. Vertical dashed lines denote the loss parameters of gold and ITO. Colored markers highlight the values of $P_\mathrm{rad}$ and $\chi$ for these materials. All calculations assume a 10 nm thick cell.}
    \label{fig:fig4}
\end{figure}

\par{So far, we have established a lossless Drude metal as the ideal substrate material for near-field TPV systems. However, the role of material loss warrants closer examination. To evaluate its impact, we vary the damping rate $\gamma_\mathrm{sub}$ of the Drude model while keeping the normalized cell thickness fixed at $\delta = 1$, reoptimizing the remaining substrate parameters for each value of $\gamma_\mathrm{sub}$. As shown in Fig. \ref{fig:fig4}, as losses increase, the objective function, $\Psi$, decreases sharply, from over 737 Wcm$^{-2}$ for $\gamma_\mathrm{sub} = 0$ to just 36 Wcm$^{-2}$ for $\gamma_\mathrm{sub} = 1$ eV. Beyond a threshold substrate loss of approximately $\gamma_\mathrm{sub} \approx 0.01$ eV, the optimal substrate transitions from a plasmonic metal to vacuum (variations of substrate parameters with loss and dielectric functions are provided in SI), highlighting the detrimental effect of increasing substrate losses on both $P_\mathrm{rad}$ and $\chi$.}

\par{To summarize, we showed that careful selection of the substrate is a key and hitherto underutilized approach to alleviate the power-efficiency trade-off in near-field TPV systems. By tailoring the substrate’s optical response and thickness, we identify conditions that confine thermal emission into a narrow spectral range just above the cell's bandgap \en where photon absorption efficiently yields electron-hole pairs. The best-performing substrate is a thin lossless Drude layer, resonantly matched to the emitter’s SPP frequency. It unlocks more than an order-of-magnitude enhancement in radiative power density compared to conventional substrates, while simultaneously achieving high spectral efficiency through the suppression of sub-bandgap losses. Multilayer or mirror-backed substrates yield only marginal benefits, reinforcing the sufficiency of a single optimized layer. As material losses increase, the optimal substrate transitions from a plasmonic metal to vacuum, underscoring the importance of low-loss materials for sustaining high performance. These principles complement ongoing efforts to tailor emitter properties for enhanced selectivity and power output. Additional structural complexity such as 2D patterning (e.g., gratings or metamaterials) may yield further enhancements and/or less stringent constraints on material loss by enabling suppression of extracted radiation at undesirable (below-gap) frequencies through field cancellations and/or pseudogap effects \cite{joannopoulos_photonic_2011}. These results establish clear design principles for near-field spectral shaping and open new opportunities for substrate-guided optimization in TPV systems.}

\par{This work was supported by the Fundació Mir-Puig, the Fundació Cellex, the Generalitat de Catalunya (2021 SGR 01443), the Ministerio de Ciencia e Innovación (CEX2019-000910-S, PID2020-112625GB-I00, PID2021-125441OA-I00) and the European Union (CATHERINA, 101168064). Views and opinions expressed are however those of the author(s) only and do not necessarily reflect those of the European Union or European Defence Agency. Neither the European Union nor the granting authority can be held responsible for them. The authors thank Dr. Benjamin Strekha and Dr. Weiliang Jin for helpful discussions. The authors declare no competing financial interest. K.N.N. acknowledges the ICFO Mobility Grant 2023 for supporting a research stay at Princeton University.}

\section*{Supporting Information Available}
Additional derivations, optimization details, data generated from optimization, supplementary figures.

\bibliographystyle{ieeetr}
\bibliography{references}

\begin{thebibliography}{10}

\bibitem{datas_thermophotovoltaic_2021}
A.~Datas and R.~Vaillon, ``Thermophotovoltaic energy conversion,'' in {\em Ultra-{High} {Temperature} {Thermal} {Energy} {Storage}, {Transfer} and {Conversion}}, pp.~285--308, Elsevier, 2021.

\bibitem{polder_theory_1971}
D.~Polder and M.~Van~Hove, ``Theory of {Radiative} {Heat} {Transfer} between {Closely} {Spaced} {Bodies},'' {\em Phys. Rev. B}, vol.~4, pp.~3303--3314, Nov. 1971.

\bibitem{pendry_radiative_1999}
J.~B. Pendry, ``Radiative exchange of heat between nanostructures,'' {\em J. Phys.: Condens. Matter}, vol.~11, pp.~6621--6633, Sept. 1999.

\bibitem{narayanaswamy_surface_2003}
A.~Narayanaswamy and G.~Chen, ``Surface modes for near field thermophotovoltaics,'' {\em Appl. Phys. Lett.}, vol.~82, pp.~3544--3546, May 2003.

\bibitem{laroche_near-field_2006}
M.~Laroche, R.~Carminati, and J.-J. Greffet, ``Near-field thermophotovoltaic energy conversion,'' {\em Journal of Applied Physics}, vol.~100, p.~063704, Sept. 2006.

\bibitem{shen_surface_2009}
S.~Shen, A.~Narayanaswamy, and G.~Chen, ``Surface {Phonon} {Polaritons} {Mediated} {Energy} {Transfer} between {Nanoscale} {Gaps},'' {\em Nano Lett.}, vol.~9, pp.~2909--2913, Aug. 2009.

\bibitem{mittapally_near-field_2021}
R.~Mittapally, B.~Lee, L.~Zhu, A.~Reihani, J.~W. Lim, D.~Fan, S.~R. Forrest, P.~Reddy, and E.~Meyhofer, ``Near-field thermophotovoltaics for efficient heat to electricity conversion at high power density,'' {\em Nat Commun}, vol.~12, p.~4364, July 2021.

\bibitem{lucchesi_near-field_2021}
C.~Lucchesi, D.~Cakiroglu, J.-P. Perez, T.~Taliercio, E.~Tournié, P.-O. Chapuis, and R.~Vaillon, ``Near-{Field} {Thermophotovoltaic} {Conversion} with {High} {Electrical} {Power} {Density} and {Cell} {Efficiency} above 14\%,'' {\em Nano Lett.}, vol.~21, pp.~4524--4529, June 2021.

\bibitem{giteau_thermodynamic_2023}
M.~Giteau, M.~F. Picardi, and G.~T. Papadakis, ``Thermodynamic performance bounds for radiative heat engines,'' {\em Phys. Rev. Applied}, vol.~20, p.~L061003, Dec. 2023.

\bibitem{nefzaoui_selective_2012}
E.~Nefzaoui, J.~Drevillon, and K.~Joulain, ``Selective emitters design and optimization for thermophotovoltaic applications,'' {\em Journal of Applied Physics}, vol.~111, p.~084316, Apr. 2012.

\bibitem{xiao_ultrabroadband_2025}
C.~Xiao, M.~Liu, K.~Yao, Y.~Zhang, M.~Zhang, M.~Yan, Y.~Sun, X.~Liu, X.~Cui, T.~Fan, C.~Zhao, W.~Hua, Y.~Ying, Y.~Zheng, D.~Zhang, C.-W. Qiu, and H.~Zhou, ``Ultrabroadband and band-selective thermal meta-emitters by machine learning,'' {\em Nature}, vol.~643, pp.~80--88, July 2025.
\newblock Publisher: Springer Science and Business Media LLC.

\bibitem{datas_optimum_2015}
A.~Datas, ``Optimum semiconductor bandgaps in single junction and multijunction thermophotovoltaic converters,'' {\em Solar Energy Materials and Solar Cells}, vol.~134, pp.~275--290, Mar. 2015.

\bibitem{king_fundamental_2025}
R.~R. King, ``Fundamental advantages of multijunction thermophotovoltaic cells,'' {\em Solar Energy Materials and Solar Cells}, vol.~293, p.~113780, Dec. 2025.
\newblock Publisher: Elsevier BV.

\bibitem{nimje_hot-carrier_2024}
K.~N. Nimje, M.~Giteau, and G.~T. Papadakis, ``Hot-carrier thermophotovoltaic systems,'' {\em J. Opt.}, vol.~26, p.~075902, July 2024.

\bibitem{omair_ultraefficient_2019}
Z.~Omair, G.~Scranton, L.~M. Pazos-Outón, T.~P. Xiao, M.~A. Steiner, V.~Ganapati, P.~F. Peterson, J.~Holzrichter, H.~Atwater, and E.~Yablonovitch, ``Ultraefficient thermophotovoltaic power conversion by band-edge spectral filtering,'' {\em Proc. Natl. Acad. Sci. U.S.A.}, vol.~116, pp.~15356--15361, July 2019.

\bibitem{fan_near-perfect_2020}
D.~Fan, T.~Burger, S.~McSherry, B.~Lee, A.~Lenert, and S.~R. Forrest, ``Near-perfect photon utilization in an air-bridge thermophotovoltaic cell,'' {\em Nature}, vol.~586, pp.~237--241, Oct. 2020.

\bibitem{lapotin_thermophotovoltaic_2022}
A.~LaPotin, K.~L. Schulte, M.~A. Steiner, K.~Buznitsky, C.~C. Kelsall, D.~J. Friedman, E.~J. Tervo, R.~M. France, M.~R. Young, A.~Rohskopf, S.~Verma, E.~N. Wang, and A.~Henry, ``Thermophotovoltaic efficiency of 40\%,'' {\em Nature}, vol.~604, pp.~287--291, Apr. 2022.

\bibitem{karalis_squeezing_2016}
A.~Karalis and J.~D. Joannopoulos, ``‘{Squeezing}’ near-field thermal emission for ultra-efficient high-power thermophotovoltaic conversion,'' {\em Sci Rep}, vol.~6, p.~28472, July 2016.

\bibitem{inoue_near-field_2018}
T.~Inoue, K.~Watanabe, T.~Asano, and S.~Noda, ``Near-field thermophotovoltaic energy conversion using an intermediate transparent substrate,'' {\em Opt. Express}, vol.~26, p.~A192, Jan. 2018.
\newblock Publisher: Optica Publishing Group.

\bibitem{bright_performance_2014}
T.~J. Bright, L.~P. Wang, and Z.~M. Zhang, ``Performance of {Near}-{Field} {Thermophotovoltaic} {Cells} {Enhanced} {With} a {Backside} {Reflector},'' {\em Journal of Heat Transfer}, vol.~136, p.~062701, June 2014.

\bibitem{inoue_near-field_2021}
T.~Inoue, T.~Suzuki, K.~Ikeda, T.~Asano, and S.~Noda, ``Near-field thermophotovotaic devices with surrounding non-contact reflectors for efficient photon recycling,'' {\em Opt. Express}, vol.~29, p.~11133, Mar. 2021.

\bibitem{feng_improved_2022}
D.~Feng, S.~K. Yee, and Z.~M. Zhang, ``Improved performance of a near-field thermophotovoltaic device by a back gapped reflector,'' {\em Solar Energy Materials and Solar Cells}, vol.~237, p.~111562, Apr. 2022.

\bibitem{lee_air-bridge_2022}
B.~Lee, R.~Lentz, T.~Burger, B.~Roy-Layinde, J.~Lim, R.~M. Zhu, D.~Fan, A.~Lenert, and S.~R. Forrest, ``Air-{Bridge} {Si} {Thermophotovoltaic} {Cell} with {High} {Photon} {Utilization},'' {\em ACS Energy Lett.}, vol.~7, pp.~2388--2392, July 2022.

\bibitem{roy-layinde_sustaining_2022}
B.~Roy-Layinde, T.~Burger, D.~Fan, B.~Lee, S.~McSherry, S.~R. Forrest, and A.~Lenert, ``Sustaining efficiency at elevated power densities in {InGaAs} airbridge thermophotovoltaic cells,'' {\em Solar Energy Materials and Solar Cells}, vol.~236, p.~111523, Mar. 2022.

\bibitem{lim_enhanced_2023}
J.~Lim, B.~Roy-Layinde, B.~Liu, A.~Lenert, and S.~R. Forrest, ``Enhanced {Photon} {Utilization} in {Single} {Cavity} {Mode} {Air}-{Bridge} {Thermophotovoltaic} {Cells},'' {\em ACS Energy Lett.}, vol.~8, pp.~2935--2939, July 2023.

\bibitem{zhao_high-performance_2017}
B.~Zhao, K.~Chen, S.~Buddhiraju, G.~Bhatt, M.~Lipson, and S.~Fan, ``High-performance near-field thermophotovoltaics for waste heat recovery,'' {\em Nano Energy}, vol.~41, pp.~344--350, Nov. 2017.

\bibitem{fiorino_nanogap_2018}
A.~Fiorino, L.~Zhu, D.~Thompson, R.~Mittapally, P.~Reddy, and E.~Meyhofer, ``Nanogap near-field thermophotovoltaics,'' {\em Nature Nanotech}, vol.~13, pp.~806--811, Sept. 2018.

\bibitem{ilic_overcoming_2012}
O.~Ilic, M.~Jablan, J.~D. Joannopoulos, I.~Celanovic, and M.~Soljačić, ``Overcoming the black body limit in plasmonic and graphene near-field thermophotovoltaic systems,'' {\em Opt. Express}, vol.~20, p.~A366, May 2012.

\bibitem{burger_present_2020}
T.~Burger, C.~Sempere, B.~Roy-Layinde, and A.~Lenert, ``Present {Efficiencies} and {Future} {Opportunities} in {Thermophotovoltaics},'' {\em Joule}, vol.~4, pp.~1660--1680, Aug. 2020.

\bibitem{song_near-field_2015}
B.~Song, A.~Fiorino, E.~Meyhofer, and P.~Reddy, ``Near-field radiative thermal transport: {From} theory to experiment,'' {\em AIP Advances}, vol.~5, p.~053503, May 2015.

\bibitem{papadakis_thermodynamics_2021}
G.~T. Papadakis, M.~Orenstein, E.~Yablonovitch, and S.~Fan, ``Thermodynamics of {Light} {Management} in {Near}-{Field} {Thermophotovoltaics},'' {\em Phys. Rev. Applied}, vol.~16, p.~064063, Dec. 2021.

\bibitem{francoeur_near-field_2008}
M.~Francoeur, M.~P. Mengüç, and R.~Vaillon, ``Near-field radiative heat transfer enhancement via surface phonon polaritons coupling in thin films,'' {\em Applied Physics Letters}, vol.~93, p.~043109, July 2008.

\bibitem{jin_overcoming_2017}
W.~Jin, R.~Messina, and A.~W. Rodriguez, ``Overcoming limits to near-field radiative heat transfer in uniform planar media through multilayer optimization,'' {\em Opt. Express}, vol.~25, p.~14746, June 2017.

\bibitem{zhang_optimal_2020}
L.~Zhang and O.~D. Miller, ``Optimal {Materials} for {Maximum} {Large}-{Area} {Near}-{Field} {Radiative} {Heat} {Transfer},'' {\em ACS Photonics}, vol.~7, pp.~3116--3129, Nov. 2020.

\bibitem{caldwell_low-loss_2015}
J.~D. Caldwell, L.~Lindsay, V.~Giannini, I.~Vurgaftman, T.~L. Reinecke, S.~A. Maier, and O.~J. Glembocki, ``Low-loss, infrared and terahertz nanophotonics using surface phonon polaritons,'' {\em Nanophotonics}, vol.~4, pp.~44--68, Apr. 2015.

\bibitem{maier_plasmonics_2007}
S.~A. Maier, {\em Plasmonics: fundamentals and applications}.
\newblock New York: Springer, 2007.

\bibitem{pascale_tight_2023}
M.~Pascale and G.~T. Papadakis, ``Tight {Bounds} and the {Role} of {Optical} {Loss} in {Polariton}-{Mediated} {Near}-{Field} {Heat} {Transfer},'' {\em Phys. Rev. Applied}, vol.~19, p.~034013, Mar. 2023.

\bibitem{economou_surface_1969}
E.~N. Economou, ``Surface {Plasmons} in {Thin} {Films},'' {\em Phys. Rev.}, vol.~182, pp.~539--554, June 1969.
\newblock Publisher: American Physical Society (APS).

\bibitem{mcsherry_effects_2019}
S.~McSherry, T.~Burger, and A.~Lenert, ``Effects of narrowband transport on near-field and far-field thermophotonic conversion,'' {\em J. Photon. Energy}, vol.~9, p.~1, Mar. 2019.
\newblock Publisher: SPIE-Intl Soc Optical Eng.

\bibitem{papadakis_broadening_2020}
G.~T. Papadakis, S.~Buddhiraju, Z.~Zhao, B.~Zhao, and S.~Fan, ``Broadening near-field emission for performance enhancement in thermophotovoltaics,'' {\em Nano Lett.}, vol.~20, pp.~1654--1661, Mar. 2020.
\newblock arXiv:1909.08701 [physics].

\bibitem{joannopoulos_photonic_2011}
J.~D. Joannopoulos, J.~N. Winn, and S.~G. Johnson, {\em Photonic {Crystals}: {Molding} the {Flow} of {Light} - {Second} {Edition}}.
\newblock Princeton, NJ: Princeton University Press, 2011.

\end{thebibliography}

\end{document}